# Evaluation and determination of seven and five parameters of a photovoltaic generator by an iterative method

Ahmed YAHFDHOU[1, 2], Abdel Kader MAHMOUD[2] and Issakha YOUM1[1, 3]

[1]Laborotory of Semiconductors and Solar Energy (LASES), Faculty of Sciences and Technology, University of Cheikh Anta Diop of Dakar_Senegal.
[2]Applied Reserch Center of Renewble Energy (CRAER), Faculty of Sciences and Technology, University of Sciences Technology and Medecine of Nouakchott_Mauritania
[3]Center of the Study and Reseach of the Renewable energy (CERER), Dakar_Senegal

**Abstract:**
*The mathematical modeling of solar cells is essential for any optimization operation of the efficiency or the diagnostics of the photovoltaic generator. The photovoltaic module is generally represented by an equivalent circuit whose parameters are experimentally calculated by using the characteristic current-tension, I-V. The precise determination of these parameters stays a challenge for the researchers, what led to a big diversification in the models and the digital methods dedicated to their characterizations. In the present paper; we are interested in the parametric characterization of a model in both following cases: with single and two diodes, in order to plan the behavior of the photovoltaic generator under real functioning conditions. We developed an identification method of the parameters using Newton Raphson's method by using the software Matlab/Simulink. This method is a fast technique which allows the identification of several parameters and can be used in real time applications. The results of the proposed method show a high agreement between the experimental and simulated characteristics photovoltaic generator.*

### I. Introduction

Electrical energy needs are still increasing over these last years but production constraints like pollution [1] and global warming [2] lead to development of renewable energy sources, particularly photovoltaic energy [3].

To surmount the problem of modeling of solar panels will have to us had a precise knowledge of the parameters of a cell PV essentially for the conception, the quality control and for the evaluation of their implementation. These parameters are often determined from experimental data for well determined climatologically conditions [4, 5].

The behavior of a module is usually described by his characteristic current-tension (I-V), the look of which depends considerably on values of the parameters such as the current of saturation, the photocurrent, serial resistance, shunt and of the dark current [6, 7].

### II. Models of the photovoltaic array

The solar panel is the main source of energy of the whole system PV. He's establishes a set of serial associated photovoltaic cells and in parallels with ports of additional protections. To precede to its analysis the basic idea will be based on the study of the photovoltaic cell, the hard core of the converter of the light in electricity. Several models of photovoltaic cells are proposed in the literature revues [8, 10, 11], among these mathematical models the models in a diode is generally retained as the most adapted to model of solar cell in normal functioning.

### II.1.     Single diode model

This photovoltaic cell is characterized by its equivalent plan (figure 1) which consists of a source of electric current which models the conversion of the luminous flow in electrical energy; a diode models the junction of the cell. To take into account physical phenomena at the level of the cell, the model is completed by two resistances series $R\_S$ and shunt $R\_{Sh}$ as the watch the equivalent electric plan. The serial resistance is due to the contribution of the basic resistances and the front of the junction and the contacts face before and back. The resistance shunt result from losses by recombination's owed essentially to the thickness; it is to reduce following penetration of the metallic impurities in the junction. The equivalent circuit of a photovoltaic cell is represented by the figure (1).

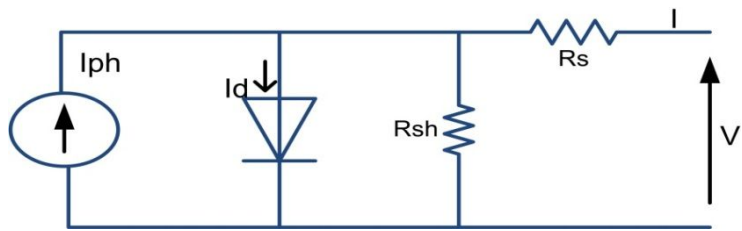

*Figure 1 : Conventional single diode model.*

This model can be expressed by the following equation, in which I and V are respectively current and the tension of a cell PV [9].

$$I = I_{ph} - Is\left(e^{\frac{V+R_s I}{a V_t}} - 1\right) - \frac{V + R_s I}{R_{sh}} \quad (1)$$

It is a not linear equation to two unknowns (I and V) and five parameters to be determined. These parameters are:

- Iph: photo-current, proportional equivalent current in the period of sunshine received by the cell.
- Is: is the reverse saturation current.
- a: is the diode ideality factor.
- $R_S$: is the equivalent series Resistance.
- $R_{SH}$: is the parallel Resistance

### II.2.     Two diodes Model

We have, this time two diodes to present the phenomena of polarization of the junction p-n. These diodes symbolize the recombination of the minority carriers, on one hand on-surface of the material and on the other hand in the volume of the material. The plan of the photovoltaic generator becomes in this case that of the figure (2):

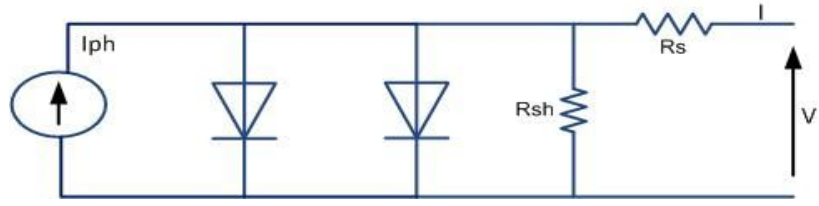

*Figure 2: equivalent electric model of solar cells, two diodes.*

Characteristic I-V of the model has two diode is to describe by this equation:

$$I = I_{ph} - I_{s1}\left[\exp\left(\frac{V+IR_s}{a_1 V_{t1}}\right)-1\right] - I_{s2}\left[\exp\left(\frac{V+IR_s}{a_2 V_{t2}}\right)-1\right] - \left(\frac{V+IR_s}{R_{sh}}\right) \quad (2)$$

Where Is1 and Is2 is the reverse saturation currents of the first diode and second diode respectively, $V_{t1}$ and $V_{t2}$ are the thermal voltages respective diodes. $a_1$ and $a_2$ represent the diode ideality constants.

### II.3. Determination of the parameters
#### II.3.1 Determination of single diode model parameters:

The equation 1 is an implicit equation there I and V which can be solved thanks to the method of Newton-Raphson or analytically with approaches, this equation has five parameters Iph, Is, has, Rs and Rsh are determined.

The current $I_{PH}$ of photovoltaic cells varies according to its temperature, the period of sunshine which receives and the coefficient of temperature at the short current circuit. By making informed reference $I_{PH\_REF}$ measured on the standard test condition ($G_{REF}$ = 1000 W / m² and $T_{REF}$=25°C). The current $I_{PH}$ for a period of sunshine and a temperature given can be calculated by the following expression:

$$I_{ph} = \frac{G}{G_{REF}}\left[I_{PH,REF} + \alpha(T-T_{REF})\right] \quad (3)$$

Where:
G: Solar irradiance received by the module area [W/m²].
α: Temperature coefficient of the short circuit current [A/K].
$G_{REF}$: Solar irradiance at the Standard Test Condition [1000W/m²].
$T_{REF}$: temperature at the Standard Test Condition [25 °C].
$I_{PH, REF}$: Light-generated at the reference condition, is practically equal of short circuit current at the STC, $I_{CC, REF}$.
The expression of the diode reverse saturation current is given by:

$$I_s = \frac{I_{cc,ref} + \alpha(T-T_{ref})}{\exp\left[(V_{oc,ref} + \beta(T-T_{ref}))/aV_t\right]-1} \quad (4)$$

Where $V_{oc, ref}$ open circuit voltage at the reference condition, ß is the temperature coefficient of open circuit voltage and $I_{cc, ref}$ is the short circuit current at standard test condition. Constants α and ß are supplied by the manufacturer of the photovoltaic cell. The serial resistances and parallel are initialized by the following equations

$$R_{s0} = -\left(\frac{dV}{dI}\right)_{V=V_{OC}} \quad (5)$$

$$R_{sh0} = -\left(\frac{dV}{dI}\right)_{I=I_{CC}} \quad (6)$$

### II.3. 2 Determination of two diodes model parameters:

This model is model mathematically described by the equation (2) which it expresses according to six unknown parameters. In this paper we consider that both running of saturation are equal to facilitate calculates him of running, we can write:

$$I_{s1} = I_{s2} = I_s = \frac{I_{ph,ref} + \alpha(T - T_{ref})}{\exp[(V_{oc,ref} + \beta(T - T_{ref}))/\{(a_1 + a_2)/p\}V_t] - 1} \quad (7)$$

This approach is to introduce by [12] by resting (a1+a2)/p=1. The resistance shunt can express by this relation:

$$R_{sh} = \frac{V_m + I_m R_s}{\left\{I_{ph} - I_s\left[\exp\left(\frac{V_m + I_m R_s}{V_t}\right) + \left(\frac{V_m + I_m R_s}{(p-1)V_t}\right) - 2\right] - \frac{P_{max}}{V_m}\right\}} \quad (8)$$

Vm and Im the voltage and the current at maximum power point, Pmax is power at maximum power point delivered by the module, and these data generally are supplied by the manufacturer data sheets (Table 1). The initial value of the resistance shunt is can express by this formula:

$$R_{sh0} = \left(\frac{V_m}{I_{cc} - I_m}\right) - \left(\frac{V_{oc} - V_m}{I_m}\right) \quad (9)$$

In this case we can will choose the initial value of the serial resistance equal zero.

*Table 1: Electrical data of the ATERSA solar module at 25°C, 1000W/m² by the manufacturer.*

| DESIGNATIONS | ATERSA AP-7105/A-75A |
| --- | --- |
| Maximum Power ($P_{max}$) | 75W |
| Voltage at Maximum Power ($V_{mp}$) | 17V |
| Current at Maximum Power ($I_{mp}$) | 4.4A |
| Open Circuit Voltage ($V_{oc}$) | 21V |
| Short Circuit Current ($I_{sc}$) | 4.8A |
| Number of cells ($N_S$) | 36 |

## II.4. Iterative method of extraction of the parameters PV

In this paper we use Newton Raphson's method to determine the parameters of two models but also to solve the equations 1 and 2. This method is described by this expression [13]:

$$X_{n+1} = X_n - \frac{f(X_n)}{f'(X_n)} \quad (10)$$

Such as f' is by-product of the function f (x) 0, $x_n$ is the present value and $x_{n+1}$ is the next value. The algorithm based on the methods of Newton Raphson for the determinations of the photovoltaic parameters is presented to the following figure (3):

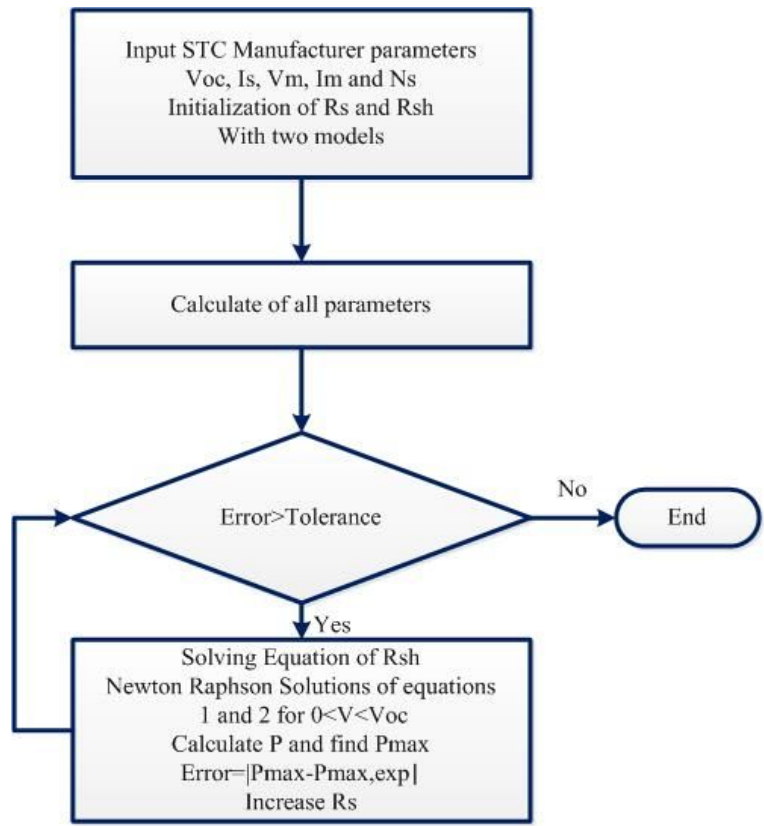

*Figure 3: Matching algorithm (Newton Raphson Method).*

## III. Experimental System Description

The experimental device consists of a hybrid system (figure. 4) of electricity production (photovoltaic, wind, diesel and storage system) of power 5,7kW coupled with a desalination plant of brackish waters (reverse osmosis) and other equipment's. The installation of the CRAER consists of following elements:

- A photovoltaic generator compound of 16 panels ATERSA (AP-7105 / AP-75), delivers a power of 1.2KWc.
- Two wind generators produce a power of 3KW.

- A system of electrochemical storage compound of 24 batteries.
- Diesel Group.
- A unity of data acquisitions such as the climatologically data and the parameters of the system.

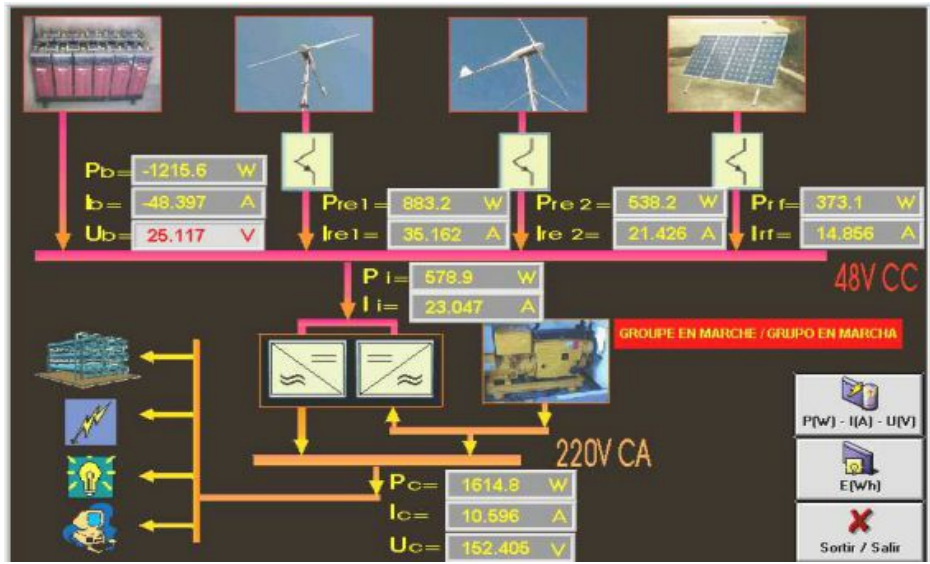

*Figure 4: The hybrid system of the Applied Center of Renewable Energy in Mauritania.*

## IV. Results and discussions

The improved iterative method technique is proposed to identify the parameters of PV module in this section. The efficiency of the improved NRM-based parameters identification method is verified by identifying the experimental data of PV module under different irradiance and temperature conditions.

Comparisons with the optimization algorithm for identification are also presented for the experimental data, which are generated using the PV module models. This algorithm is programmed and implemented in MATLAB environment to identify the PV module parameters using the two models of the photovoltaic module, the parameters identified by this implementation of the algorithm are presented to the table (2).

*Table 2: Parameters for simulated by iterative method.*

| Parameters | Single diode model | Two diode model |
|---|---|---|
| $I_{ph}$ | 4.81 | 4.8 |
| $I_{s1}$ | $9.965 \times 10^{-10}$ | $6.620 \times 10^{-10}$ |
| $I_{s2}$ | - | $6.620 \times 10^{-10}$ |
| $R_s$ | 0.28 | 0.27 |
| $R_{sh}$ | 115.9 | 112.42 |
| $a_1$ | 1.02 | 1 |
| $a_2$ | - | 1.2 |

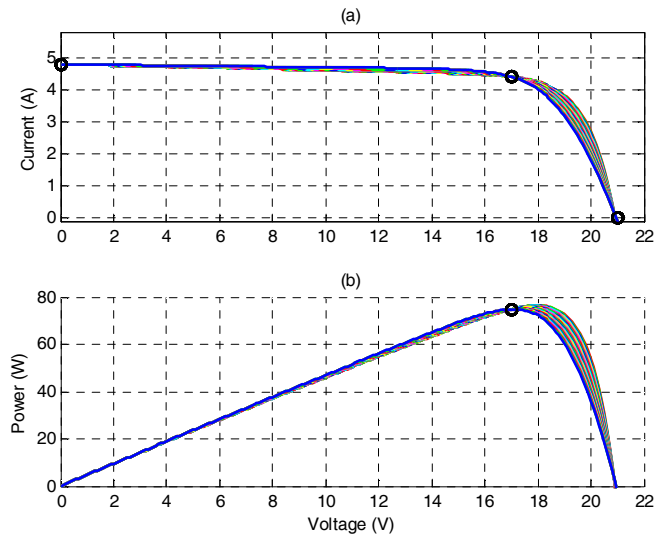

*Figure 5: Adjusting I-V (a) and P-V (b) curves for different varies of Rs and Rsh (Single diode model) at the STC.*

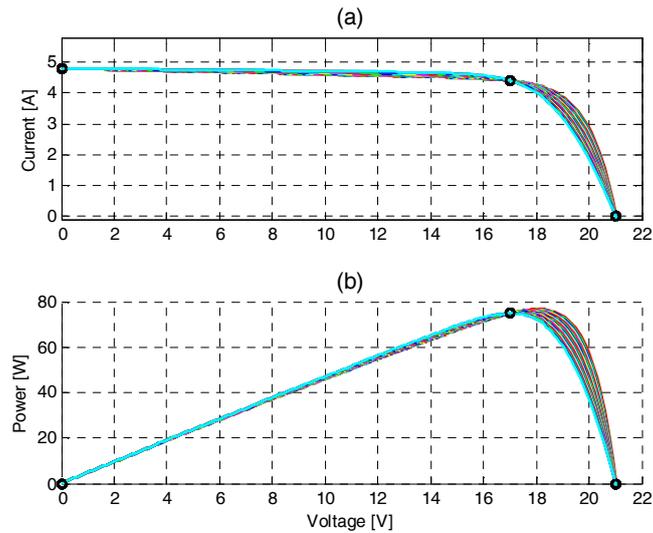

*Figure 6: Adjusting I-V (a) and P-V (b) curves for different varies of Rs and Rsh (Single diode model) at the STC.*

Figs. 5 and 6 show the I-V and P-V curves of the ATERSA photovoltaic array adjusted with the NRM. The curves exactly match with the experimental data at the three remarkable points provided by the datasheet: short current circuit, maximum power, and open circuit voltage. Table (1) shows the experimental parameters of the array obtained from the datasheet.

This method for adjusting Rs and Rsh based on the fact that there is an only pair (Rs, Rsh) that warranties that *Pmax,m = Pmax,e = Vm*Im* at the (Vm, Im) point of the I–V curve, the maximum power calculated by the model (Pmax,m) is equal to the maximum experimental power from the datasheet (Pmax,e) at the MPP.

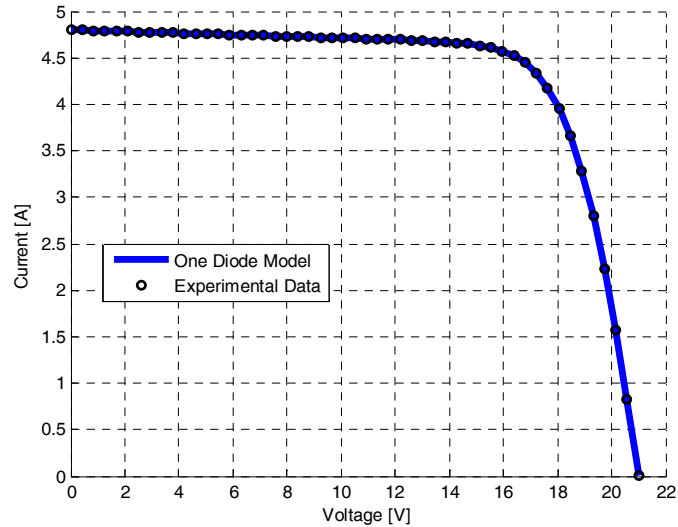

*Figure 7: A comparison of V-I curve using the identified parameters from the NRM (solid line) and the experimental data (dots) for ATERSA (single diode).*

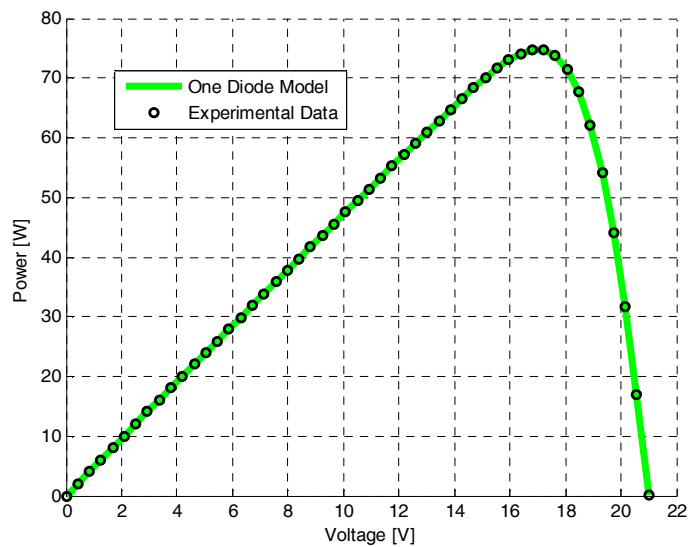

*Figure 8: A comparison of V-P curve using the identified parameters from the NRM (solid line) and the experimental data (dots) for ATERSA (single diode).*

Figures, 7 and 8 represent the V–I and V-P curves for single diode model under the standard test conditions of G = 1000 W/m2, T = 298.2 K. Based on these curves, the necessary specifications were calculated (Table 2).

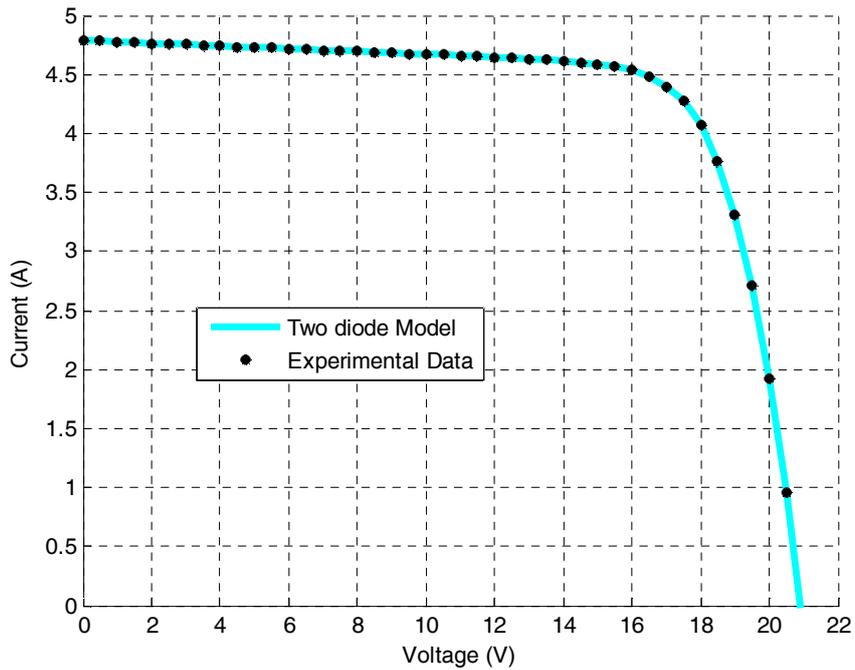

*Figure 9: A comparison of V-I curve using the identified parameters from the NRM (Solid line) and the experimental data (dots) for ATERSA (two diodes model).*

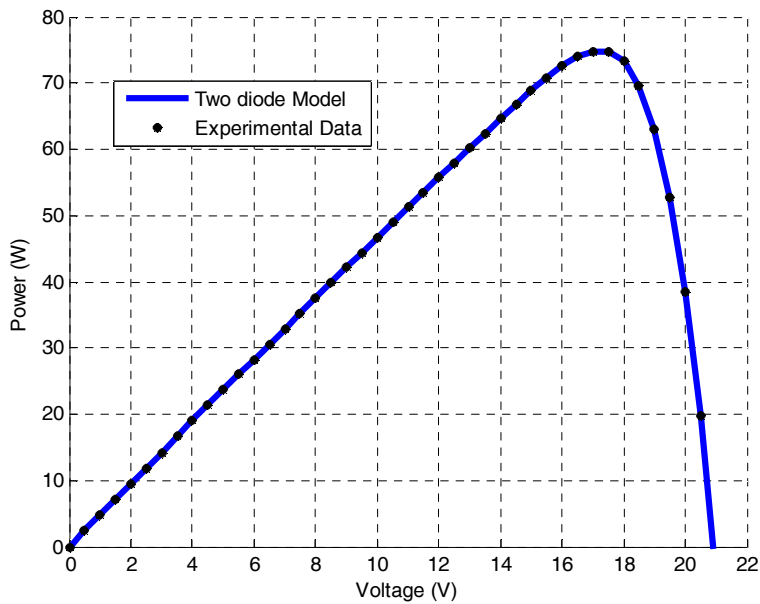

*Figure 10: A comparison of V-P curve using the identified parameters from the NRM (solid line) and the experimental data (dots) for ATERSA (Two diodes model).*

Figure 9 and 10, have V-I and V-P curves for the dual diode model according to the standard test conditions, we find that there is a very good correlation between the model and experimental datasheet.

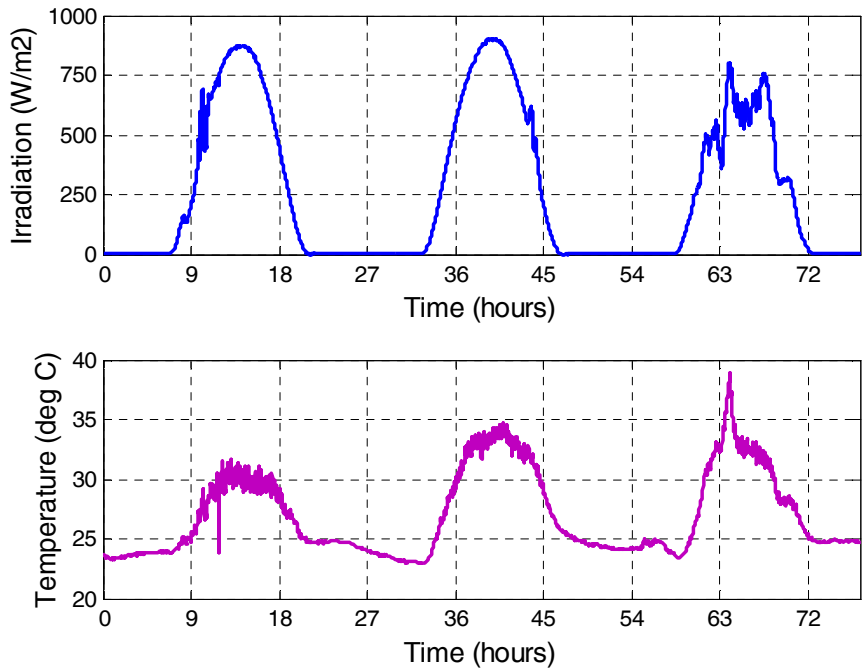

*Figure 11: The sunshine (a) and temperature (b) profiles recorded for three days at experimental site research center applied to renewable energy.*

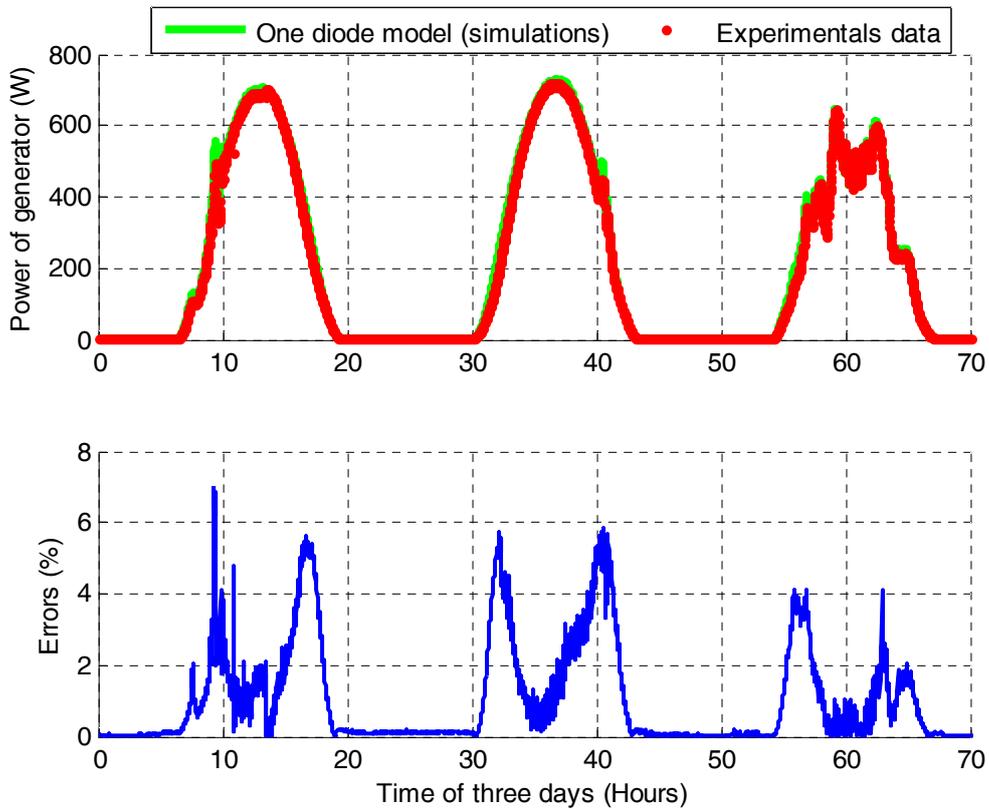

*Figure 12: comparison between the simulated power and experimental raised power (a) for three days in the case of single diode model, (b) relative error.*

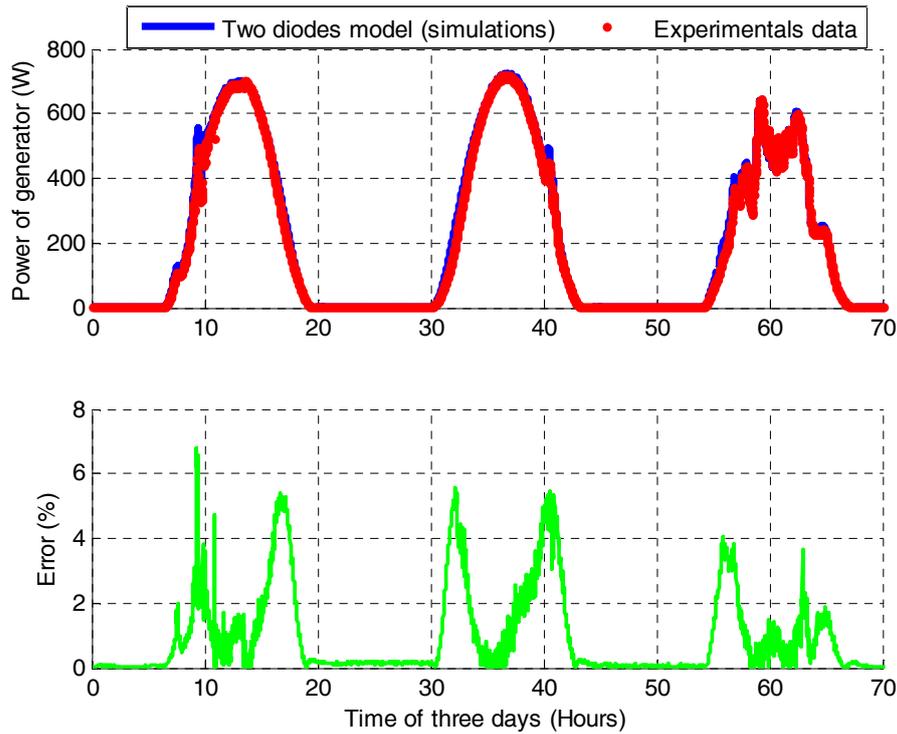

*Figure 1 3: comparison between the simulated power and experimental raised power (a) for three days in the case of single diode model, (b)relative error.*

*Table 3: Relative power of error for both models*

| Models | Maximum relative error (%) | | | Mean RE (%) |
|---|---|---|---|---|
| | Fist day | Second day | Third day | Three days |
| Single diode model | 7.014 | 7.014 | 4.089 | 1.0260 |
| Two diodes model | 6.818 | 5.574 | 4.037 | 0.9264 |

The figure 11 represents the profile of sunlight and temperature noted at applied research center for renewable energy for three days of the month of August. We use the weather data for the validation of models based on the extraction of five and seven parameters of the PV generator, follows simulate the power delivered by the PV system during these test days.

Meteorological data show that it will be a remarkable perturbation on the third day caused by the transit of clouds. The maximum irradiation saved to quarter past eleven does not exceed 800 W / m² while the first day of the order of eight hundred eighty and for the second day slightly exceeds nine hundred.

The variation of temperature during the three days shows that the maximum temperature is observed in the third day around 11 and a half of about 39 ° C although the 1st and the second day do not exceed respectively 32 and 35 degrees Celsius.

The result in Figure 12 show a comparison between the power estimate by single diode model and the experience of our generator that consists of 16 identical panels, as well as the relative error between simulated and real power during three days of tests within our site.

These results obtained shows a very good correlation between simulated values and experimental data has real records using the data acquisition system during the period of testing of the photovoltaic system. The relative error calculation shows that by simulating the power model is very close to thse actual power delivered by the PV system despite the remarkable changes in weather conditions of the site.

Figure 13 shows a comparison of the power obtained by the model of two diodes and power of real or experimental GPV three days of the test, and also the variation of relative error between these powers.

The power generated by the relative error two models bitter Table 3 shows that the model of two diode is near the most adapt to the behavior continues the photovoltaic system.

## V. Conclusion

In this paper, the seven and five parameters of a solar PV module are extracted using an iterative method based Newton Raphson Method (NRM).The feasibility of the NR method has been validated by synthetic and experimental I–V and P-V data set of solar module (mono-crystalline). It was found that proposed method is very accurate and converges to the solution very rapidly. The results also presented good agreement between modeling data and outdoor measurements of generator photovoltaic for single and two diode models, even under low irradiance levels. The mean absolute relative error between estimated power and measurements is less than 1.1 % for single diode model parameters and less than 1% for two diode model.

## References

[1] Askarzadeh, A. Razazadeh, Extraction of maximum power point in solar cells using bird mating optimizer-based parameters identification approach, Solar Energy 90, pp. 123-133, 2013.
[2] M. R. Alrashidi, M. F. Alhajri, K.M. El-naggar, A. K. Al-othman, A new estimation approach for determining the I-V characteristics of solar cells, Solar Energy 85, pp. 1543-1550, 2011.
[3] M. Seifi, A. B. Chesoh, N. I. Abdwahab, M.KB. Hasan, A comparative study of PV models in Matlab/Simulink, Word Academy of Science, Engineering and Technology 74, pp. 108-113, 2013.
[4] A. N. Celik, N. Acikgoz, Modeling and experimental verification of the operating current of mono-crystalline photovoltaic modules using four and five parameter models, Applied Energy 84, 1-5, 2007.
[5] A. A. Zadeh and A. Rezazadeh, Artificial bee swarm optimization algorithm for parameters identifications of solar cell modules, Applied Energy 102, 943-949, 2013/
[6] L. Sandrolini, M. Artioli, U. Reggiani, Numerical method for extraction of photovoltaic modules double-diode model parameters through cluster analysis, Applied Energy 87, 442-451, 2010.
[7] F. Dakichi, B. Ouakarfi, A. Fakkar, N. Belbounaguia, Parameters identification of solar cell model using Levenberg-Marquardt algorithm combined with simulated annealing, Solar energy 110, 781-788, 2014.
[8] D. Bonkoungou, Z. Koalaga, D. Njomo, Modeling and simulation of photovoltaic module considering single-diode equivalent circuit model in Matlab, International Journal of Emerging Technology and Advanced Engineering 3(3), pp. 493-502, 2013.
[9] A.Yahfdhou, A.Mahmoud, I.Youm, Modeling and optimization of photovoltaic generator with Matlab/Simulink, International Journal of I Tech and E Engineering 3(4), pp. 108-111, 2013.
[10] M. Yahya, I. Youm, A. Kader, Behavior and performance of a photovoltaic generator in real time, International Journal of the Physical Science 6(18),pp. 4361-4367, 2011.
[11] Ishaque K, Salam Z, Syafaruddin. A comprehensive MATLAB Simulink PV system simulator with partial shading capability based on two-diode model. Sol Energy 2011; 85:2217–27.
[12] K. Ishaque, Z. Salam, H. Thateri, Simple, fast and accurate two diode model for photovoltaic modules. Solar Energy Materials & solar Cells 95, 586-594, 2011.
[13] D. Bonkoungou, Z. Koalaga, D. Njomo, Modeling and simulation of photovoltaic module considering single-diode equivalent circuit model in Matlab. International Journal of Emerging Technology and Advanced Engineering 3(3), pp. 493-502, 2013.